\documentclass{article}

\usepackage[utf8]{inputenc} 
\usepackage[T1]{fontenc}    
\usepackage{hyperref}       
\usepackage{url}            
\usepackage{booktabs}       
\usepackage{amsfonts}       
\usepackage{nicefrac}       
\usepackage{microtype}      
\usepackage{graphicx}
\usepackage{subcaption}
\usepackage[table,xcdraw]{xcolor}
\PassOptionsToPackage{options}{natbib}

\usepackage[margin=1.25in]{geometry}

\title{Expressive Neural Voice Cloning}

\author{ \textbf{*}Paarth Neekhara, \textbf{*}Shehzeen Hussain,\\ Shlomo Dubnov, Farinaz Koushanfar, Julian McAuley\\
University of California San Diego\\
\textbf{*}Equal contribution\\
}

\begin{document}

\maketitle

\begin{abstract}
Voice cloning is the task of learning to synthesize the voice of an unseen speaker from a few samples. While current voice cloning methods achieve promising results in Text-to-Speech (TTS) synthesis for a new voice, these approaches lack the ability to control the expressiveness of synthesized audio.
In this work, we propose a controllable voice cloning method that allows fine-grained control over various style aspects of the synthesized speech for an unseen speaker. We achieve this by explicitly conditioning the speech synthesis model on a speaker encoding, pitch contour and latent style tokens during training. 
Through both quantitative and qualitative evaluations, we show that our framework can be used for various expressive voice cloning tasks using only a few transcribed or untranscribed speech samples for a new speaker.
These cloning tasks include style transfer from a reference speech, synthesizing speech directly from text, and fine-grained style control by manipulating the style conditioning variables during inference.
\footnote{Audio examples: \url{https://expressivecloning.github.io/}\\Interactive Demo: \url{https://expressivecloning.github.io/app.html}}
\end{abstract}

\section{Introduction}

Recent research efforts in voice cloning have focused on synthesizing a person's voice from only a few reference audio samples. While such a system can generate speech from text for a new speaker, it leaves out control over various style aspects of speech.
Explicit control over the style aspects of cloned speech is desirable for several applications, such as: voice-overs in animated films, synthesizing realistic and expressive speech for DeepFake videos, translating speech from one language to another while preserving speaking style and speaker identity, advertisement campaigns with expressive speech in multiple voices and languages (etc.). 
Expressive voice cloning systems can also help create personalized speech interfaces with voice assistants in smartphones, cars, and home assistants. 
Since speech serves as a primary communication interface between machine learning agents and humans, the ability to speak \textit{expressively} is a very desirable quality for voice cloning systems.
Furthermore, such systems can potentially empower individuals who have lost their ability to speak.

The goal of voice cloning is commonly formulated as learning to synthesize the voice of an unseen speaker using only a few seconds of transcribed or untranscribed speech.
This is typically done by embedding speaker-dependent information from the available speech samples of the new speaker, and conditioning a trained multi-speaker Text-to-Speech (TTS) model on the derived speaker embedding~\cite{neuralvoicecloning,transferspeakerverification}. 
While such a system can achieve promising results in closely retaining speaker-specific characteristics in the cloned speech, it does not offer control over other aspects of speech that are not contained in the text or the speaker-specific embedding. 
These aspects include variation in tone, speaking rate, emphasis and emotions. 

Several past works have focused on the problem of expressive TTS synthesis by learning latent variables for controlling the style aspects of speech synthesized for a given text~\cite{gst,prosody}. Such models are usually trained on a single-speaker expressive speech dataset to learn meaningful latent codes for various style aspects of the speech.
Recent works~\cite{tpgst,mellotron}, have extended the idea of learning style representations to a multi-speaker setting by conditioning the TTS synthesis model on both speaker identity and style encodings. 
Such techniques show promise in disentangling style and speaker specific information, and generate different style variants of synthesized speech for the same text and speaker. However, these methods are limited by the speakers used in the training set and cannot be directly used for synthesizing voices of speakers not seen during training.

Adapting multi-speaker TTS models for voice cloning requires scaling up model training to a large multi-speaker TTS dataset, containing several minutes of transcribed speech from thousands of speakers. High speaker diversity in the training data is important to achieve generalization on unseen speakers~\cite{neuralvoicecloning,transferspeakerverification}. 
\textit{The goal of our work is to perform TTS synthesis for an unseen speaker with control over the style aspects of generated speech}. As a first step in this direction, we train a TTS model conditioned on speaker encodings and latent style tokens~\cite{gst} on a large multi-speaker dataset. 
While this model is able to generate voices for unseen speakers, we find that the results fall short in terms of speech naturalness and style control during synthesis. Our results suggest that learning meaningful latent style aspects is difficult when training on a large multi-speaker dataset containing speech with mostly neutral style and expressions.

To address problem of disentangling style and speaker characteristics on a large multi-speaker dataset containing mostly style-neutral speech, we propose a voice cloning model that is conditioned on both latent and heuristically derived style information. Specifically, we condition our TTS synthesis model on (i) text, (ii) speaker encoding (iii) pitch contour of the target speech and (iv) latent style tokens~\cite{gst}. By conditioning synthesis on various style aspects and speaker embeddings derived from the target speech, we are able to train a model that offers fine-grained style control for synthesized speech. To adapt inference for an unseen speaker, we can either perform zero-shot inference or fine-tune the synthesis model on the limited text and speech pairs for the new speaker. Through both quantitative and qualitative evaluations, we demonstrate that our proposed model can make a new voice express, emote, sing or copy the style of a given reference speech. 

\section{Background and Related Work}

\textbf{Neural TTS:} State-of-the-art neural approaches for natural TTS synthesis~\cite{ping2017deep,shen2018natural} typically decompose the waveform synthesis pipeline into two steps: (1) Synthesizing perceptually informed mel-spectrograms from language using an attention based sequence-to-sequence model like Tacotron~\cite{tacotron} or Tacotron 2~\cite{shen2018natural}. (2) Vocoding the synthesized spectrograms to audible waveforms using a neural vocoder~\cite{wavenet,waveglow,advoc} or heuristic methods like the Griffin-Lim~\cite{griffinlim} algorithm. Multi-speaker TTS models~\cite{gibiansky2017deep,Ping2017DeepV3} extend this line of work by additionally conditioning the spectrogram synthesis model on speaker embeddings, which are trained end-to-end using the speaker labels in the TTS dataset. While these approaches achieve promising results in synthesizing speech for multiple speakers for a given text, they cannot be directly used to synthesize voices of speakers not seen during training.

\textbf{Voice Cloning:} Voice cloning focuses on generative modeling of speech conditioned on a speaker encoding derived from a few reference speaker audio samples. While speech synthesis models exist~\cite{wavenet,tacotron}, it has been a challenge to adapt these voice models to new speakers with limited data. 
Recent efforts have been made in designing systems that can learn to synthesize a person’s voice from only a few audio samples~\cite{neuralvoicecloning,chen2019cross,icaspVC,9053104,transferspeakerverification}. They train a separate speaker encoding network to condition a multi-speaker TTS model on speaker dependent information. Since the speaker encoding network operates on waveforms, it can be used for zero-shot voice cloning from untranscribed utterances of a target speaker. Additionally, the authors of~\cite{neuralvoicecloning} demonstrate that the synthesis model can be fine-tuned on limited text and audio pairs of a new speaker to improve the speaker similarity of the cloned speech.

\textbf{Expressive Speech Synthesis:} Prior works~\cite{gst,tpgst,prosody} on expressive speech synthesis focus on models that can be conditioned on text and a latent embedding for style or prosody. During training, the style embeddings are derived using a learnable module called \textit{Global Style Tokens (GST)}, that operates on the target speech for a given phrase and derives a style embedding through attention over a dictionary of learnable vectors. During inference, the synthesizer can be conditioned on different reference audios to produce style variants of speech for the same text. Manipulating these latent style variables during inference offers some coarse control over the style of the synthesized speech. 
Recently proposed Mellotron model~\cite{mellotron} uses a combination of explicit and latent style variables to offer more fine-grained control over the expressive characteristics of synthesized speech. Specifically, Mellotron conditions the spectrogram synthesis network on pitch contour, GSTs~\cite{gst} and speaker ID during training. During inference, the synthesizer can be conditioned on the melodic information---pitch and rhythm of a reference speech and synthesize speech in the voice of a given speaker in the training set. The authors demonstrate that explicit conditioning on pitch contour during training phase, makes it possible to generalize the inference on various melodic pitch contours. 

\section{Methodology}

Our expressive voice cloning framework is a multi-speaker TTS model that is conditioned on speaker encodings and style aspects of speech. Style conditioning in expressive TTS models is popularly done by learning a dictionary of latent style vectors called Global Style Tokens (GST)~\cite{gst}.
While GSTs can learn meaningful latent codes when trained on a dataset with high variation in expressions, we empirically find that it offers limited style control when trained on a large multi-speaker dataset with mostly neutral prosody.

\begin{figure}
    \centering
    \includegraphics[width=0.8\linewidth]{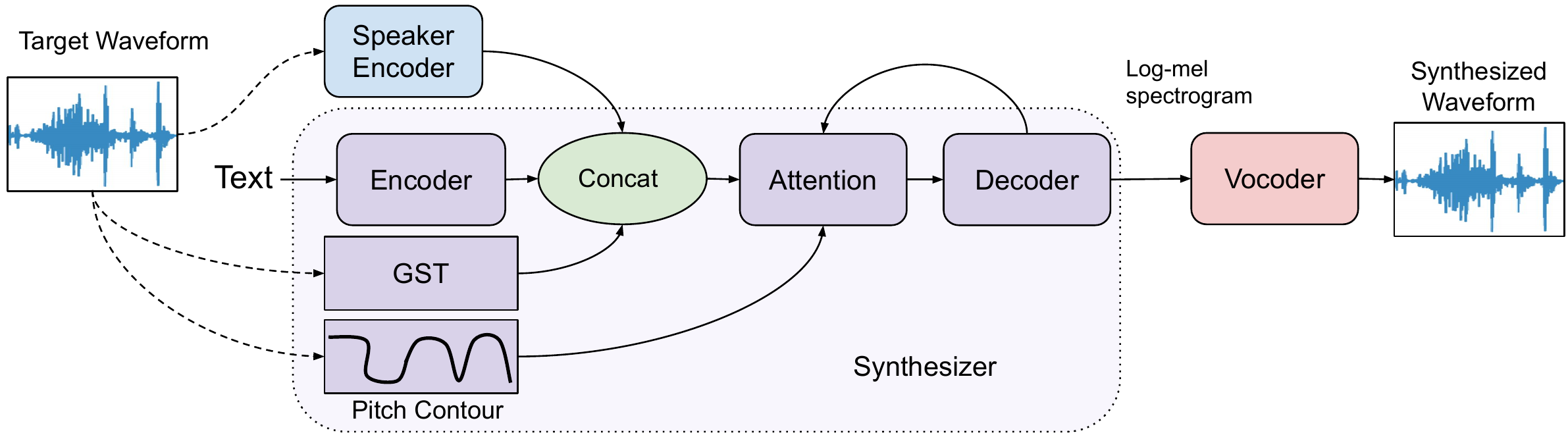}
    \caption{Expressive Voice Cloning Model: Tacotron-2 TTS model conditioned on speaker and style characteristics derived from the target audio of a given text. At inference time, the model can be provided independent references for style and speaker encodings to achieve expressive voice cloning.}
    \label{fig:model_diagram}
    \vspace{-4mm}
\end{figure}

Signal processing heuristics like the Yin algorithm~\cite{de2002yin} can derive the fundamental frequency contour (pitch contour) and voicing decisions from speech, which can be useful for expressive speech synthesis. We find that using a combination of latent and heuristically derived style information in the TTS model not only provides fine-grained control over the style aspects of synthesized speech, but also scales up to a large multi-speaker dataset to produce more natural sounding audio for an unseen speaker.
A high level overview of our expressive voice cloning framework is shown in Figure \ref{fig:model_diagram}. Similar to past works on voice cloning~\cite{neuralvoicecloning,transferspeakerverification}, the three main components \emph{Speaker Encoder}, \emph{Mel Spectrogram Synthesizer} and \emph{Vocoder} are all trained separately. 
We describe the individual components of our framework and their training objectives in the following sections.  

\subsection{Speaker Encoder}
\label{sec:speakerencoder}
Speaker conditioning in multi-speaker TTS models is usually done using a lookup in the speaker embedding matrix which is randomly initialized and trained end-to-end with the synthesizer. While such a framework learns speaker-specific information via the embedding vectors, synthesis cannot be generalized to unseen speakers. To adapt the multi-speaker TTS model for the goal of voice cloning, the speaker embedding layer can be replaced with a speaker encoder that derives speaker specific information from the target waveform. In this setting, the speaker encoder can obtain embeddings for speakers not seen during training using a few reference speech samples. To obtain meaningful embeddings, the speaker encoder should be trained to discriminate between different speakers for the task of speaker verification~\cite{generalizedspeakerverification}. 

We 
follow the speaker encoder architecture described in~\cite{generalizedspeakerverification,resemblyzer}. The network is a stack of 3 LSTM layers with 256 cells in each layer that operate on mel-spectrograms with 40 channels. The final speaker embedding is obtained by projecting the LSTM output at the last layer to 256 dimensions followed by $L_2$ normalization. Note that ours is a smaller model than that used in~\cite{transferspeakerverification} which had 768 cells in each LSTM layer. 
The speaker encoder is trained to optimize a generalized end-to-end speaker verification loss~\cite{generalizedspeakerverification}, that encourages high cosine similarity between embeddings from same speaker and low similarity between different speaker embeddings. 
During inference, each utterance is broken into smaller segments of 1,600 ms with 1,000 ms overlap between consecutive segments. The final embedding is estimated by averaging the embedding of each individual segment. 


\subsection{Mel-Spectrogram Synthesizer}
\label{sec:melsynthesizer}
The goal of our synthesis model is to disentangle the style and speaker-specific information in speech by conditioning our TTS synthesis model on the speaker encoding and various style aspects. To this end, we adapt the synthesis model used in Mellotron~\cite{mellotron} for the task of voice cloning. Mellotron is a multi-speaker TTS model that extends Tacotron 2 GST~\cite{gst} by additional conditioning on pitch contours and speaker embeddings. To adapt Mellotron for voice cloning, we remove the speaker embedding layer and replace it with the speaker encoder network described in Section~\ref{sec:speakerencoder}. 

At its core, our synthesis model based on Tacotron 2~\cite{shen2018natural}, is an LSTM based sequence-to-sequence model composed of an encoder that operates on a sequence of characters and a decoder that generates the individual frames of the mel spectrogram while attending over the encoded representations. Along with the encoded representation for text, we concatenate the speaker encoding (obtained from the speaker encoder) and the GST embedding at each time-step. The GST embedding is obtained by querying a dictionary of latent style vectors with the target mel-spectrogram using a multi-headed attention mechanism described in~\cite{gst}. 
Decoding occurs in an autoregressive manner where we synthesize one mel spectrogram frame at a time by providing the fundamental frequency (from the pitch contour) and the mel spectrogram of the previous frame as the input to the decoder. The pitch contours are derived from the target speech using the Yin algorithm with harmonicity thresholds between 0.1 and 0.25.

In this way, we can factor mel-spectrogram synthesis into the following variables: 
\emph{text ($t$)}, \emph{speaker encoding ($s$)}, \emph{pitch contour ($f_0$)} and \emph{latent style embedding obtained from GST ($z$)}. Formally, our synthesizer is a generative model $g(t, s, f_0, z; W)$ that is parameterized by trainable weights $W$, trained to optimize a loss function $L$ that penalizes the differences between the generated and ground truth mel spectrogram. That is,
\begin{equation}
\min_{W}~ 
    \mathbb{E}_{ (t_i, {a}_i) \sim D } \left \{ L(g(t_i, s_i, {f_0}_i, z_i; W), \mathit{mel}_i) \right \}
    \label{eq:synthesis}
\end{equation}
where $D$ is the dataset containing text and audio pairs $(t_i, {a}_i)$. The variables $(s_i, {f_0}_i, z_i, \mathit{mel}_i)$ are all derived from the target waveform $a_i$. For the loss function $L$, we use the L2 loss between the generated and ground truth mel spectrograms.

During training, the synthesizer learns another latent variable: the attention map between the encoder and decoder states which captures the alignment between text and audio. Following the notation used in~\cite{mellotron}, we call this latent variable \emph{rhythm}, since it controls the timing aspects of synthesized speech.  Note that unlike other style aspects which can be obtained directly from $a_i$, deriving \emph{rhythm} requires both text and audio $(t_i, {a}_i)$. In our experiments, we obtain the \emph{rhythm} by using our synthesizer as a forced-aligner. That is, for a given text and audio pair, we derive the attention map between the encoder and decoder states by doing a forward pass through our model using teacher forcing. Therefore, during inference, our synthesizer $g$ can be explicitly conditioned on rhythm $r$ derived from some text and audio pair: $g(t, s, f_0, z, \textbf{\textit{r}}; W)$. 

While the style aspects are obtained from the target waveform of the same speaker during training, we can use a different reference audio and text pair during inference. For example, we can transfer the pitch contour and rhythm of a style reference audio $S$ from a different speaker to the voice of a given target speaker $T$ as follows:
\begin{equation}
\mathit{mel} = g(t_{S}, s_{T}, {f_0}_{S}, z_{T}, r_S; W)
\end{equation}
The output $\mathit{mel}$ should have the same pitch and rhythm as the style reference $S$ and should retain the latent style aspects and voice of the target speaker $T$. In our work we focus on three different cloning tasks with different sources of style conditioning information which we discuss in Section \ref{sec:cloningtasks}.


Additionally, to assess the importance of pitch contours during training, 
we train another TTS model that is conditioned only on the latent style aspects obtained using GST. We use the same Tacotron2 architecture and GST module as our proposed model. 
Formally, this alternative synthesizer $g(t, s, z; W)$ is trained to optimize the same objective as Equation \ref{eq:synthesis}:
\begin{equation}
\min_{W}~ 
    \mathbb{E}_{ (t_i, {a}_i) \sim D } \left \{ L(g(t_i, s_i, z_i; W), \mathit{mel}_i) \right \}
    \label{eq:taco2GST}
\end{equation}
We refer to this alternative model as \textit{Tacotron2 + GST} in our experiments.
Similar to our proposed system, this model can also be additionally conditioned on rhythm. 
Since we are not explicitly conditioning the model on pitch contours, we expect the pitch variation in speech to be captured as part of the latent style tokens.
We empirically demonstrate that using only latent style representation on a large multi-speaker dataset with neutral prosody offers limited style control and audio naturalness.

\textbf{Vocoder:} For decoding the synthesized mel-spectrograms into listenable waveforms, we use the WaveGlow~\cite{waveglow} model trained on the single speaker Sally dataset~\cite{mellotron}. An advantage of WaveGlow over WaveNet~\cite{wavenet} is that it allows real-time inference, while being competitive in terms of audio naturalness. 
The same vocoder model is used across all experiments and datasets. We find that the vocoder model trained on a single speaker generalizes well across all speakers in our datasets.

\subsection{Cloning Techniques: Zero-Shot and Model Adaptation}
We adopt the following two approaches for cloning the voice of a new speaker from a few transcribed or untranscribed speech samples:

\noindent \textbf{Zero-Shot:} 
For zero-shot voice cloning, we derive the speaker embedding by taking the mean followed by L-2 normalization of the speaker encodings of the individual samples of the target speaker. 
Since speaker encodings are obtained directly from the waveforms, we do not require audio transcriptions of the new speaker for zero-shot voice cloning.

\noindent \textbf{Model Adaptation:}
When transcribed samples of a new speaker are available, we can fine-tune our synthesis model using the text and audio pairs. 
As shown in Neural Voice Cloning~\cite{neuralvoicecloning}, fine-tuning can significantly improve the speaker similarity metrics of the cloned speech. Also, the authors of~\cite{neuralvoicecloning} observe that fine-tuning the whole synthesis model is faster and more effective than fine-tuning only the speaker embedding layer since more degrees of freedom are allowed in the whole model adaptation. Our preliminary experiments on model adaptation suggested the same. We hypothesize the reason for this is that fine-tuning the last-few layers of the synthesis model is essential, if not sufficient, to adapt the synthesizer to the speaker-specific speech characteristics. Therefore, we study the following two model adaptation techniques:
\textbf{Adaptation whole -} Fine-tune all the parameters of the synthesis model on the text and audio pairs of the new speaker. 
\textbf{Adaptation decoder -} Fine-tune only the decoder parameters of the synthesis model. The advantage of only adapting the decoder parameters is that it requires fewer speaker-specific model parameters and a shared encoder can be used across all speakers in a real-world deployment setting. In both of the above adaptation settings, we fine-tune our model for 100 to 200 iterations using Adam optimizer with a learning rate of 1e-4. Model adaptation takes up to 6 minutes for fine-tuning on 1 to 20 samples of the target speaker on a single Nvidia Titan 1080 GPU.

\section{Experiments}
\subsection{Datasets and Training}
We train our mel-spectrogram synthesis model on the clean subset of the publicly available Libri-TTS~\cite{librittsref} dataset---\textit{train-clean-100} and \textit{train-clean-360}. This clean subset contains around 245 hours of speech across 1151 speakers sampled at 24 kHz. Past works on voice cloning~\cite{generalizedspeakerverification,neuralvoicecloning} trained their synthesis models on the LibriSpeech dataset~\cite{librispeechref} and empirically demonstrated the importance of a speaker-diverse training dataset for the task of voice cloning. 
We filter out utterances longer than 10 seconds and resample waveforms to 22050 Hz.

For training the synthesizer, we warm start our model using the pre-trained Mellotron checkpoint which is trained on a subset of LibriTTS containing 123 speakers. 
The speaker embedding layer is replaced with our speaker encoding network which is kept frozen during training. We use a validation set with 250 examples and train the model using a batch size of 32 and an initial learning rate of 5e-4. We use an Adam optimizer~\cite{kingma2014adam} to update the weights and anneal the learning rate to half its value every 50k mini-batch iterations. We point to our codebase~\footnote{Codebase to be relased upon publication} for precise model implementation. 
For the \textit{Tacotron 2 + GST} model, we use the same Tacotron 2 architecture and GST hyper-parameters as our proposed model. Training for the proposed model and the \textit{Tacotron 2 + GST} model converged in 210,000 and 185,000 mini-batch iterations respectively and took around 4 seconds per iteration on a single Nvidia Titan 1080 GPU. The Resemblyzer speaker encoder~\cite{resemblyzer,realtime} used in our experiments is trained on the VoxCeleb~\cite{Nagrani19}, VoxCeleb2~\cite{Chung18b} and LibriSpeech-other~\cite{librispeechref} datasets containing a total of 8.4k speakers. The authors of~\cite{realtime} report a 4.5\% Equal Error Rate (EER) for the task of speaker verification using this speaker encoder on their internal test set.

\subsection{Cloning Tasks}
\label{sec:cloningtasks}
In this section, we discuss the three main tasks for which we evaluate our voice cloning methods. When cloning the voice of a new speaker, we require a few audio samples of the speaker to obtain the speaker encoding. We refer to these samples as \textit{target speaker samples}. We perform voice cloning for the speakers in the VCTK dataset~\cite{vctk}. The VCTK dataset contains speech sampled at 48 KHz from 108 native English speakers, the majority of which have British accents. We down-sampled the audio to 22,050 KHz to make it consistent with our training data. To synthesize the speech for a given speaker encoding and text, our synthesis model additionally requires various style conditioning variables described in Section \ref{sec:melsynthesizer}. While the latent GST embedding can be obtained from the \textit{target speaker samples}, pitch contour and rhythm information needs to be derived from a \textit{style reference audio} that corresponds to the given text. In case we do not have a style reference audio available, we can synthesize one using a single speaker TTS system. 
To evaluate our cloning techniques objectively in terms of style and speaker disentanglement, and also assess their usefulness in real world settings, we perform the following cloning tasks:

\noindent\textbf{1. Text} \textit{Cloning speech directly from text: }
For cloning speech directly from text, we first synthesize speech for the given text using a single speaker TTS model: Tacotron 2 + WaveGlow trained on the LJ Speech~\cite{ljspeech17} dataset. We then derive the pitch contour of the synthetic speech using the Yin algorithm~\cite{de2002yin} and scale the pitch contour linearly to have the same mean pitch as that of the \textit{target speaker samples}. For deriving rhythm, we use our proposed synthesis model as a forced aligner between the text and Tacotron2-synthesized speech. We use the \textit{target speaker samples} for obtaining the GST embedding for both our proposed model and the baseline Tacotron2 + GST model.

\noindent\textbf{2. Imitation -} \textit{Reconstruct a sample from the target speaker:} In this setup, we use a text and audio pair of the target speaker (not contained in the \textit{target speaker samples}), and try to reconstruct the audio from its factorized representation using our synthesis model. All of the style conditioning variables - pitch, rhythm and GST embedding are derived from the speech sample we are trying to imitate. The imitation task is a toy experiment that allows quantitative evaluation of style similarity metrics between the synthesized speech and style reference.
\\
\noindent\textbf{3. Style Transfer -} \textit{Transfer the pitch and rhythm of speech from a different expressive speaker:} The goal of this task is to transfer the pitch and rhythm from some expressive speech to the cloned speech for the target speaker. For this task, we use examples from the single speaker Blizzard 2013 dataset~\cite{King2013TheBC} as style references. This dataset contains expressive audio book readings from a single speaker with high variation in emotion and pitch.
For our proposed model, we use this \textit{style reference audio} to extract the pitch and rhythm. Similar to the Text task, we scale the pitch contour to have the same mean as that of the \textit{target speaker samples}. In-order to retain speaker-specific latent style aspects, we use \textit{target speaker samples} to extract the GST embedding. 
For the Tacotron2 + GST model, which does not have explicit pitch conditioning, we use the \textit{style reference audio} for obtaining the GST embedding and the rhythm.

\subsection{Results}
For the above described cloning tasks, we evaluate three aspects of the cloned speech: i) speaker similarity to the target speaker, ii) style similarity to the reference style and iii) speech naturalness. We encourage the readers to listen to our audio examples referenced in the footnote of the first page to contextualize the following results.

\begin{figure}[h]
    \centering
    \includegraphics[width=1.\linewidth]{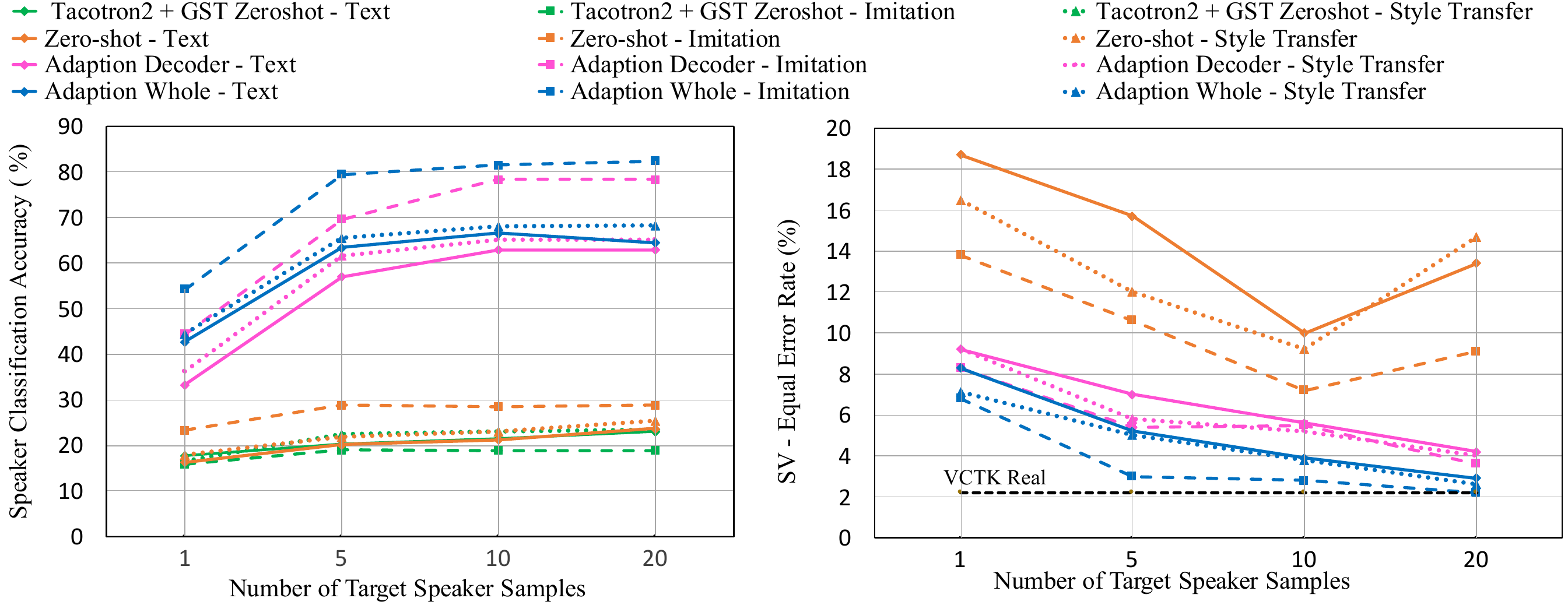}
    \caption{Speaker similarity evaluation of each cloning technique for different voice cloning tasks in terms of Speaker Classification Accuracy and Speaker Verification Equal Error Rate (SV-EER).}
    \label{fig:speakergraph}
\end{figure}

\noindent \textbf{Speaker Classification Accuracy:} We train a speaker classifier on the VCTK dataset to classify a given utterance as one of the 108 speakers. The speaker classifier is a two layer neural network with 256 hidden units that takes as input the speaker encoding obtained through our pre-trained speaker encoder network.
Similar to ~\cite{neuralvoicecloning}, our speaker classifier achieves 100\% accuracy on a hold out set containing 200 examples from the VCTK dataset. However, since our classification model and training dataset for the synthesizer are not the same as~\cite{neuralvoicecloning} (1,151 speakers in ours vs.~2,481 speakers in~\cite{neuralvoicecloning}), we do not make direct comparisons with their work.
We conduct our speaker classification evaluations on all 108 speakers of the VCTK dataset. We clone 25 speech samples per speaker for each task described in Section \ref{sec:cloningtasks}. 
Figure \ref{fig:speakergraph} (left) shows the speaker classification accuracy curves for all cloning tasks and techniques with respect to the number of target speaker samples. 
Our results are 
consistent
with the following findings of~\cite{neuralvoicecloning}---Model adaptation significantly outperforms the zero-shot voice cloning technique since it allows the model to adjust to the speaker characteristics of the new speaker. More 
target speaker samples helps improve speaker classification accuracy, although in the zero-shot scenario we do not observe much improvement after 10 target speaker samples.

For zero-shot voice cloning, both Tacotron2-GST and our proposed model achieve similar speaker classification accuracy for \textit{Text} and \textit{Style Transfer} cloning tasks. The accuracy of our proposed model is slightly higher for the imitation task as compared to other tasks for both model adaptation and zero-shot voice cloning. This implies that conditioning on the actual pitch contour of the target speaker improves speaker specific characteristics of the cloned speech. While linear scaling of a reference style pitch contour works well, our findings motivate future research on predicting speaker-specific pitch contours from text and speaker encodings. 

\noindent \textbf{Speaker verification Equal Error Rate (SV-EER):}
SV-EER is another objective metric used to evaluate speaker similarity between the cloned audio and the ground-truth reference audio. We use a speaker verification system that scores the speaker similarity between two utterances based on the cosine similarity of the encodings obtained using the speaker encoder described in Section~\ref{sec:speakerencoder}. 
Equal Error Rate (EER) is the point when the false acceptance rate and false rejection rate of the speaker verification system are equal.

We perform speaker verification evaluations on randomly selected 20 speakers in the VCTK dataset. We enroll 5 speech samples per speaker in the speaker verification system and synthesize 50 speech samples per speaker for each cloning task. EERs are estimated by pairing each sample of the same speaker with another sample from a different speaker. Figure \ref{fig:speakergraph} shows the plots of SV-EER for different cloning techniques and tasks using our proposed model and also the those estimated using real data.
Our observations on the SV-EER metric are similar to those on the accuracy metric. Model adaptation outperforms zero-shot cloning techniques and with more than 10 cloning samples achieves similar EER as the real data. Additionally, we include human evaluation scores on speaker similarity in our supplementary material.

\begin{table}[h]
\centering
\footnotesize
\begin{tabular}{l|ccc|c}
\multicolumn{1}{c}{} & \multicolumn{3}{c}{\emph{Imitation}} & \multicolumn{1}{c}{\emph{Style Transfer}}\\
\toprule
Approach & GPE & VDE  & FFE & Style-MOS\\
\midrule
Tacotron2 + GST - Zero-shot & 20.37\% & 26.39\% & 29.47\% & $2.69 \pm 0.11$ \\
Proposed Model - Zero-shot   & 3.72\% & 10.65\% & 11.74\% & $3.15 \pm 0.11$\\
Proposed Model - Adaptation Whole  & 2.97\% & 12.58\% & 13.60\% & $3.40 \pm 0.10$\\
Proposed Model - Adaptation Decoder  & 2.39\% & 11.60\% & 12.51\% & $3.29 \pm 0.10$\\
\bottomrule
\end{tabular}

\caption{Style similarity evaluations for the imitation and style transfer tasks. We use three objective error metrics (lower values are better). For the style transfer task we present the mean opinion scores on style similarity (Style-MOS) with 95\% confidence interval.
}
\label{tab:stylesimilarity}
\end{table}




\textbf{Style Similarity:}
In order to evaluate the similarity between the style of synthesized and reference audio, we perform quantitative evaluation on the Imitation task described in Section~\ref{sec:cloningtasks}. We use the following error metrics: Gross Pitch Error (GPE)~\cite{nakatani2008method}, Voicing Decision Error (VDE)~\cite{nakatani2008method} and F0 Frame Error (FFE)~\cite{chu2009reducing}. Results are presented in Table~\ref{tab:stylesimilarity} in which we compare the error values for different approaches when using 10 target speaker samples for cloning. We synthesize 25 speech samples per speaker for all speakers in the VCTK dataset to estimate the reported error values. Our proposed models significantly outperform the Tacotron 2 + GST baseline, clearly indicating the importance of pitch contour conditioning for accurate style transfer. 

We also conduct a crowd-sourced listening test on Amazon Mechanical Turk (AMT) for the style transfer task in which we ask the listeners to rate the style similarity between the ground truth style reference and synthesized audio on a 5 point scale (interface for this study is included in the supplementary material). For each cloning technique (using 10 target speaker samples), we synthesize 25 audio samples per speaker for 20 speakers in the VCTK dataset leading to 500 evaluations of each technique.  We present the style similarity Mean Opinion Scores (Style-MOS) in Table \ref{tab:stylesimilarity}. It can be seen that our proposed models significantly outperform the Tacotron 2 + GST model. Model adaptation techniques perform slightly better than zero-shot method suggesting that fine-tuning improves the model predictions for an unseen speaker encoding.

\textbf{Naturalness:}
To assess speech naturalness, we conducted a crowd-sourced listening test on AMT and asked listeners to rate each audio utterance on a 5-point naturalness scale to collect Mean Opinion Scores (MOS). 
Similar to the above mentioned user study, we use 10 target speaker samples for each cloning technique. All evaluations are conducted on randomly selected 20 VCTK speakers with 25 audio samples synthesized per speaker. Each sample is rated independently by a single listener leading to 500 evaluations for each technique per cloning task. We report the MOS of Real data and audio synthesized using different cloning techniques in Table~\ref{tab:realMOS}. Our proposed model significantly outperforms the baseline Tacotron2 + GST model for both zero-shot and model adaptation techniques. This suggests that pitch contour conditioning in a multi-speaker setting helps improve speech naturalness in addition to providing higher style similarity. 
It can be seen that the naturalness is even further improved with model adaptation techniques since it allows the generative model to adjust for the unseen speaker encodings.

\begin{table}[h]
\centering
\footnotesize
\begin{tabular}{lccc}
\multicolumn{1}{c}{} & \multicolumn{3}{c}{\emph{Cloning Task}} \\
\toprule
Approach & Text & Imitation  & Style Transfer \\
\midrule
Real data VCTK & \multicolumn{3}{c}{$4.11 \pm 0.08$} \\
Real data Blizzard & \multicolumn{3}{c}{$4.07 \pm 0.08$} \\
\midrule
Tacotron2 + GST - Zero-shot & $2.67 \pm 0.10$ & $2.51 \pm 0.10$  & $3.02 \pm 0.09$\\
Proposed Model - Zero-shot   & $3.56 \pm 0.09$ & $3.54  \pm 0.10$ & $3.53 \pm 0.10$ \\
Proposed Model - Adaptation Whole  & $3.75 \pm 0.09$ & $3.71 \pm 0.09$ & $3.60 \pm 0.09$ \\
Proposed Model - Adaptation Decoder  & $3.61 \pm 0.09$ & $3.57 \pm 0.09$ & $3.45 \pm 0.09$ \\
\bottomrule
\end{tabular}
\caption{Mean Opinion Score (MOS) for speech naturalness with 95\% confidence intervals.}
\label{tab:realMOS}
\vspace{-7mm}
\end{table}



\section{Conclusion and Future Work}
In this work we introduce an expressive voice cloning and define three benchmark tasks to evaluate such systems. 
We empirically find that learning only latent style tokens is insufficient to capture expressiveness in speech when training the synthesis model on a speaker-diverse dataset with mostly neutral prosody.  
Our proposed model uses a combination of heuristically derived and latent style information, which not only offers fine-grained control over style aspects but also improves speech naturalness. We demonstrate that our proposed model can successfully extract and transfer style and speaker characteristics from unseen audio references to the synthesized speech. We recommend future works on models for predicting speaker specific pitch contours directly from style labels (like \textit{happy}, \textit{sad}, \textit{neutral} etc)\  and text to allow control over expressions of the synthesized speech when a style reference audio is not available.




\section{Broader Impact}
Speech interfaces enable hands-free operation and can assist users who are visually or physically impaired. Research into machine generation of speech is driven by the prospect of offering services where humans interact solely with machines, thereby eliminating the cost of live agents and significantly reducing the cost of providing services.
Since speech serves as a primary communication interface between machine learning agents and humans, the ability to speak expressively in a new voice can help create more personalized machine assistants. Furthermore, such systems can also empower individuals who have lost their ability to speak.

Explicit control over the style aspects of cloned speech is also desirable for several multimedia applications. These include: voice overs in animated films, synthesizing realistic and expressive speech for videos, translating speech from one language to another while preserving the speaking style and speaker identity, advertisement and political campaigns with expressive speech in multiple voices or languages, etc.

Our intent for generating expressive speech is to advance the research of synthetic audio generation, such that it can aid in the accessibility of speech interfaces and support users with speech impairments, as well as contribute to mainstream use in movies, digital storytelling and modern-day streaming services. This work provides us with an opportunity to collaborate with researchers for advancing multi-disciplinary investigation of AI techniques. 
However, any emerging technology can also be abused. Realistic voice cloning technology can be used to create voice-overs for subjects of DeepFake videos, and has the potential to be used maliciously to spread disinformation or for creating inappropriate content. Also, the technology can be abused for circumventing speech based user authentication systems in smart devices.
We seek to discourage the unethical use of our technology. Upon the release of public access to our voice cloning app, we plan to incorporate techniques to watermark the speech generated by our model. This will allow us to thwart the misuse of our technology and curb any spread of misinformation using our platform. 
It is our intention to develop Expressive Voice Cloning in a way that its potential for abuse is minimized and maximise its use as a tool for learning, education and experimentation.

\bibliographystyle{plain}
\bibliography{myref.bib}

\begin{thebibliography}{10}

\bibitem{neuralvoicecloning}
Sercan Arik, Jitong Chen, Kainan Peng, Wei Ping, and Yanqi Zhou.
\newblock Neural voice cloning with a few samples.
\newblock In {\em NeurIPS}. 2018.

\bibitem{chen2019cross}
Mengnan Chen, Minchuan Chen, Shuang Liang, Jun Ma, Lei Chen, Shaojun Wang, and
  Jing Xiao.
\newblock Cross-lingual, multi-speaker text-to-speech synthesis using neural
  speaker embedding.
\newblock In {\em INTERSPEECH}, 2019.

\bibitem{chu2009reducing}
Wei Chu and Abeer Alwan.
\newblock Reducing f0 frame error of f0 tracking algorithms under noisy
  conditions with an unvoiced/voiced classification frontend.
\newblock In {\em ICASSP}. IEEE, 2009.

\bibitem{Chung18b}
J.~S. Chung, A.~Nagrani, and A.~Zisserman.
\newblock Voxceleb2: Deep speaker recognition.
\newblock In {\em INTERSPEECH}, 2018.

\bibitem{icaspVC}
E.~{Cooper}, C.~{Lai}, Y.~{Yasuda}, F.~{Fang}, X.~{Wang}, N.~{Chen}, and
  J.~{Yamagishi}.
\newblock Zero-shot multi-speaker text-to-speech with state-of-the-art neural
  speaker embeddings.
\newblock In {\em ICASSP}, 2020.

\bibitem{de2002yin}
Alain De~Cheveign{\'e} and Hideki Kawahara.
\newblock Yin, a fundamental frequency estimator for speech and music.
\newblock 2002.

\bibitem{gibiansky2017deep}
Andrew Gibiansky, Sercan Arik, Gregory Diamos, John Miller, Kainan Peng, Wei
  Ping, Jonathan Raiman, and Yanqi Zhou.
\newblock Deep voice 2: Multi-speaker neural text-to-speech.
\newblock In {\em NIPS}, 2017.

\bibitem{griffinlim}
Daniel~W. Griffin, Jae, S.~Lim, and Senior Member.
\newblock Signal estimation from modified short-time {F}ourier transform.
\newblock {\em IEEE Trans. Acoustics, Speech and Sig. Proc}, 1984.

\bibitem{9053104}
Y.~{Huang}, L.~{He}, W.~{Wei}, W.~{Gale}, J.~{Li}, and Y.~{Gong}.
\newblock Using personalized speech synthesis and neural language generator for
  rapid speaker adaptation.
\newblock In {\em ICASSP}, 2020.

\bibitem{ljspeech17}
Keith Ito.
\newblock The lj speech dataset.
\newblock \url{https://keithito.com/LJ-Speech-Dataset/}, 2017.

\bibitem{transferspeakerverification}
Ye~Jia, Yu~Zhang, Ron Weiss, Quan Wang, Jonathan Shen, Fei Ren, zhifeng Chen,
  Patrick Nguyen, Ruoming Pang, Ignacio Lopez~Moreno, and Yonghui Wu.
\newblock Transfer learning from speaker verification to multispeaker
  text-to-speech synthesis.
\newblock In {\em NeurIPS}. 2018.

\bibitem{King2013TheBC}
Simon~J. King and Vasilis Karaiskos.
\newblock The blizzard challenge 2013.
\newblock In {\em In Blizzard Challenge Workshop}, 2013.

\bibitem{kingma2014adam}
Diederik~P Kingma and Jimmy Ba.
\newblock Adam: A method for stochastic optimization.
\newblock In {\em ICLR}, 2015.

\bibitem{realtime}
Gilles Louppe.
\newblock Master thesis : Automatic multispeaker voice cloning.
\newblock 2019.

\bibitem{resemblyzer}
Gilles Louppe.
\newblock {\em Resemblyzer - https://github.com/resemble-ai/Resemblyzer/},
  2019.

\bibitem{Nagrani19}
Arsha Nagrani, Joon~Son Chung, Weidi Xie, and Andrew Zisserman.
\newblock Voxceleb: Large-scale speaker verification in the wild.
\newblock {\em Computer Science and Language}, 2019.

\bibitem{nakatani2008method}
Tomohiro Nakatani, Shigeaki Amano, Toshio Irino, Kentaro Ishizuka, and Tadahisa
  Kondo.
\newblock A method for fundamental frequency estimation and voicing decision:
  Application to infant utterances recorded in real acoustical environments.
\newblock {\em Speech Communication}, 2008.

\bibitem{advoc}
Paarth Neekhara, Chris Donahue, Miller Puckette, Shlomo Dubnov, and Julian
  McAuley.
\newblock Expediting {TTS} synthesis with adversarial vocoding.
\newblock In {\em INTERSPEECH}, 2019.

\bibitem{librispeechref}
Vassil Panayotov, Guoguo Chen, Daniel Povey, and Sanjeev Khudanpur.
\newblock Librispeech: an asr corpus based on public domain audio books.
\newblock In {\em ICASSP}. IEEE, 2015.

\bibitem{ping2017deep}
Wei Ping, Kainan Peng, Andrew Gibiansky, Sercan~O Arik, Ajay Kannan, Sharan
  Narang, Jonathan Raiman, and John Miller.
\newblock Deep {V}oice 3: Scaling text-to-speech with convolutional sequence
  learning.
\newblock In {\em ICLR}, 2018.

\bibitem{Ping2017DeepV3}
Wei Ping, Kainan Peng, Andrew Gibiansky, Sercan~{\"O}. Arik, Ajay Kannan,
  Sharan Narang, Jonathan Raiman, and John~L. Miller.
\newblock Deep voice 3: 2000-speaker neural text-to-speech.
\newblock In {\em ICLR}, 2018.

\bibitem{waveglow}
Ryan Prenger, Rafael Valle, and Bryan Catanzaro.
\newblock Wave{G}low: {A} flow-based generative network for speech synthesis.
\newblock In {\em ICASSP}, 2018.

\bibitem{shen2018natural}
Jonathan Shen, Ruoming Pang, Ron~J Weiss, Mike Schuster, Navdeep Jaitly,
  Zongheng Yang, Zhifeng Chen, Yu~Zhang, Yuxuan Wang, Rj~Skerrv-Ryan, et~al.
\newblock Natural {TTS} synthesis by conditioning {W}ave{N}et on mel
  spectrogram predictions.
\newblock In {\em ICASSP}, 2018.

\bibitem{prosody}
R.~J. Skerry{-}Ryan, Eric Battenberg, Ying Xiao, Yuxuan Wang, Daisy Stanton,
  Joel Shor, Ron~J. Weiss, Rob Clark, and Rif~A. Saurous.
\newblock Towards end-to-end prosody transfer for expressive speech synthesis
  with tacotron.
\newblock {\em arXiv:1803.09047}, 2018.

\bibitem{tpgst}
Daisy Stanton, Yuxuan Wang, and R.~J. Skerry{-}Ryan.
\newblock Predicting expressive speaking style from text in end-to-end speech
  synthesis.
\newblock {\em arXiv:1803.09017}, 2018.

\bibitem{mellotron}
Rafael Valle, Jason Li, Ryan Prenger, and Bryan Catanzaro.
\newblock Mellotron: Multispeaker expressive voice synthesis by conditioning on
  rhythm, pitch and global style tokens.
\newblock {\em ICASSP}, 2020.

\bibitem{wavenet}
A\"{a}ron van~den Oord, Sander Dieleman, Heiga Zen, Karen Simonyan, Oriol
  Vinyals, Alex Graves, Nal Kalchbrenner, Andrew Senior, and Koray Kavukcuoglu.
\newblock Wave{N}et: A generative model for raw audio.
\newblock {\em arXiv:1609.03499}, 2016.

\bibitem{vctk}
Christophe Veaux, Junichi Yamagishi, and Kirsten Macdonald.
\newblock Cstr vctk corpus: English multi-speaker corpus for cstr voice cloning
  toolkit.
\newblock 2017.

\bibitem{generalizedspeakerverification}
Li~Wan, Quan Wang, Alan Papir, and Ignacio~Lopez Moreno.
\newblock Generalized end-to-end loss for speaker verification.
\newblock {\em arXiv:1710.10467}, 2017.

\bibitem{tacotron}
Yuxuan Wang, RJ~Skerry-Ryan, Daisy Stanton, Yonghui Wu, Ron~J Weiss, Navdeep
  Jaitly, Zongheng Yang, Ying Xiao, Zhifeng Chen, Samy Bengio, et~al.
\newblock Tacotron: Towards end-to-end speech synthesis.
\newblock In {\em INTERSPEECH}, 2017.

\bibitem{gst}
Yuxuan Wang, Daisy Stanton, Yu~Zhang, RJ~Skerry-Ryan, Eric Battenberg, Joel
  Shor, Ying Xiao, Fei Ren, Ye~Jia, and Rif~A. Saurous.
\newblock Style tokens: Unsupervised style modeling, control and transfer in
  end-to-end speech synthesis.
\newblock {\em arXiv:1803.09017}, 2018.

\bibitem{librittsref}
Heiga Zen, Viet Dang, Rob Clark, Yu~Zhang, Ron~J. Weiss, Ye~Jia, Zhifeng Chen,
  and Yonghui Wu.
\newblock {LibriTTS: A Corpus Derived from LibriSpeech for Text-to-Speech}.
\newblock In {\em INTERSPEECH}, 2019.

\end{thebibliography}
\end{document}